\documentclass[12pt]{article}
\pdfoutput=1
\usepackage[table]{xcolor}
\usepackage{amsmath,amssymb,exscale,cancel}
\usepackage{graphicx}
\usepackage{epsfig}
\usepackage{multicol}
\usepackage{color}
\usepackage{mathrsfs}
\usepackage{blindtext}
 \usepackage{fancyhdr}
\usepackage{hyperref}
\usepackage{cite}
\usepackage{mathtools}

\usepackage{floatrow}

\usepackage{graphicx}
\usepackage{sidecap}

\usepackage[latin1]{inputenc} 
\usepackage{multicol}  
 
 \usepackage{titlesec}
 
 \usepackage{rotating,slashed,xcolor,amsfonts,expdlist,charter}

\numberwithin{equation}{section}

\usepackage{xcolor}
\usepackage{sectsty}

\usepackage{mdframed}
\usepackage{titletoc}

\usepackage{filecontents}

\definecolor{secnum}{RGB}{13,151,225}
\definecolor{ptcbackground}{RGB}{212,237,252}
\definecolor{ptctitle}{RGB}{0,177,235}

\titlecontents{lsection}
  [5.8em]{\sffamily}
  {\color{secnum}\contentslabel{2.3em}\normalcolor}{}
  {\titlerule*[1000pc]{.}\contentspage\\\hspace*{-5.8em}\vspace*{5pt}%
    \color{white}\rule{\dimexpr\textwidth-15.5pt\relax}{1pt}}

\usepackage{hyperref}
\hypersetup{colorlinks,bookmarksopen,bookmarksnumbered,citecolor=rossos,
linkcolor=redy,pdfstartview=FitH,urlcolor=blus}
\usepackage{slashed}

\definecolor{blus}{cmyk}{1,1,0,0.1}
\definecolor{verdes}{cmyk}{0.99,0,0.59,0.65}
\definecolor{rossos}{cmyk}{0,1,1,0.55}
\definecolor{redy}{cmyk}{0,1,1,0.7}
\definecolor{greeny}{cmyk}{0.99,0,0.59,0.98}
\definecolor{green-go}{cmyk}{0.91,0,0.59,0.5}

\usepackage{titlesec}

\newcommand{\beq}{\begin{equation}}
\newcommand{\eeq}{\end{equation}}

\def\hhref#1{\href{http://arxiv.org/abs/#1}{arXiv:#1}} % in bibliography

\newcommand{\tmtextbf}[1]{{\bfseries{#1}}}
\newcommand{\tmtextrm}[1]{{\rmfamily{#1}}}
\newcommand{\bp}{\bar M_P}

\def\be{\begin{equation}}
\def\ee{\end{equation}}
\def\ba{\begin{array} }
\def\tt{{\tiny \times}}
\def\bac{\begin{array} {c}}
\def\bacc{\begin{array} {cc}}
\def\baccc{\begin{array} {ccc}}
\def\bacccc{\begin{array} {cccc}}
\def\ea{\end{array}}
\def\bea{\begin{eqnarray}}
\def\eea{\end{eqnarray}}

\definecolor{red}{rgb}{1,0,0}
\def\oo{\textcolor{blue}}
\def\psl{\hbox{\hbox{${p}$}}\kern-1.9mm{\hbox{${/}$}}}
\def\dsl{\hbox{\hbox{${\partial}$}}\kern-2.2mm{\hbox{${/}$}}}
\def\Dsl{\hbox{\hbox{${D}$}}\kern-2.6mm{\hbox{${/}$}}}

\newcommand{\gappeq}{{\rlap{{\raise}.5ex\text{\ensuremath{>}}}{{\lower}.5ex\text{\ensuremath{\sim}}}}}
\newcommand{\lappeq}{{\rlap{{\raise}.5ex\text{\ensuremath{<}}}{{\lower}.5ex\text{\ensuremath{\sim}}}}}

\newcommand{\I}{\tmtextrm{1{\kern}-.24em l}}

\begin{document}
\topmargin -1.0cm
\oddsidemargin 0.9cm
\evensidemargin -0.5cm

{\vspace{-1cm}}
\begin{center}

\vspace{-1cm}

 {\Huge \tmtextbf{ \color{blus} 
Gravitational Waves from Fundamental Axion Dynamics}} {\vspace{.5cm}}\\
 
\vspace{0.9cm}

{\large {\bf Anish Ghoshal$^{a,b}$}  and {\bf Alberto Salvio$^{a,c}$ }  
%\vspace{.4cm}\\
%{\large }\\
\vspace{.3cm}

{\em  

\vspace{.4cm}
${}^a$ I. N. F. N. -  Rome Tor Vergata,\\
via della Ricerca Scientifica, I-00133 Rome, Italy\\ 
\vspace{0.6cm}

${}^b$ I. N. F. N., Laboratori Nazionali di Frascati, \\ 
C.P. 13, I-00044 Frascati, Italy

\vspace{0.6cm}

${}^c$ Physics Department, University of Rome Tor Vergata, \\ 
via della Ricerca Scientifica, I-00133 Rome, Italy
%Email: alberto.salvio@roma2.infn.it

%To do 
% send it to JHEP

\vspace{2cm}
 
}}

\noindent ------------------------------------------------------------------------------------------------------------------------------
 \vspace{0.cm}
\end{center}

\begin{center}
{\bf \large Abstract}

\end{center}
A totally asymptotically free QCD axion model, where  all couplings flow to zero in the infinite energy limit, was recently formulated. A very interesting feature of this fundamental theory is the ability to predict some low-energy observables, like the masses of the extra fermions and scalars.
 Here we find and investigate a region of the parameter space where the Peccei-Quinn (PQ) symmetry  is  broken quantum mechanically through the Coleman-Weinberg mechanism. This results in an even more predictive framework: the axion sector features only two independent parameters (the PQ symmetry breaking scale and the QCD gauge coupling). 
 In particular, we show that the PQ  phase transition is strongly first order and can produce gravitational waves within the reach  of  future detectors. The predictivity of the model leads to 
 %specific characteristics 
a rigid dependence of the phase transition (like its duration and the nucleation temperature) and the gravitational wave spectrum on the PQ symmetry breaking scale and the QCD gauge coupling.

  \vspace{0.9cm}
\noindent    
\noindent ------------------------------------------------------------------------------------------------------------------------------
%

%\noindent  

\vspace{0.cm}
% \end{center}
% 
\noindent %--------------------------------------------------------------------------------------------------------------------------------

%\end{abstract}

\vspace{-.5cm}

\noindent %--------------------------------------------------------------------------------------------------------------------------------
 \newpage

\tableofcontents

\vspace{0.9cm}

\section{Introduction}

  The PQ symmetry~\cite{Peccei:1977hh}  provides an  elegant solution of the strong CP problem: it explains dynamically why the strong interactions preserve CP, while the electroweak ones break it.  This solution, regardless of its implementation, manifests itself at low energies through the axion, the pseudo-Goldstone boson associated with the breaking of this (approximate)  symmetry~\cite{Weinberg:1977ma}. Moreover, the axion is a good dark matter (DM) candidate and can actually account for the whole DM when the PQ symmetry breaking scale $f_a$ is around $10^{11}\,$GeV~\cite{Preskill:1982cy}. Even if there may be other components of DM, astrophysical observations require anyhow $f_a$ to be above the scale of $\sim 10^8\,$GeV (see Ref.~\cite{DiLuzio:2020wdo} for a recent review). This makes it difficult to test the PQ idea as current colliders are far from probing those energies. 

Cosmology, on the other hand, allows us to have a window on physical processes occurring at such high mass scales. A classic example is  the observation of the cosmic microwave background (CMB) anisotropies. Indeed, these  give us information on the inflationary dynamics, which typically occurs at energies much above those within the reach of colliders (see Refs.~\cite{Ade:2015lrj,Akrami:2018odb} for up-to-date observations).  Recently, the experimental  discovery of gravitational waves (GWs)~\cite{Abbott:2016blz}  has opened another  window on very energetic processes predicted by particle physics models. An important example is the possibility to have experimental information on the characteristics of phase transitions, if these are strongly first order (see Ref.~\cite{Maggiore:2018sht} for a textbook introduction to this topic). Moreover, the possible observation of GWs due to a first-order phase transition would be a remarkably clear signal of new physics as the finite-temperature symmetry breaking dynamics in the Standard Model (SM) is not of this type.
 
  If the PQ phase transition is strongly first order it could lead to observable GWs. But, as pointed out in Refs.~\cite{DelleRose:2019pgi,vonHarling:2019gme}, the particular features of the phase transition and the corresponding GWs depend on the specific dynamics implementing the PQ symmetry\footnote{See also Refs.~\cite{Dev:2016feu,Croon:2019iuh,Dev:2019njv,Machado:2019xuc} for other studies of GWs in axion and axion-like effective models.}. These features include, for example, the temperature at which the phase transition occurs, its duration and the GW spectrum. 

One way one can significantly reduce this ambiguity, and thus have a guide regarding the directions towards which the experimental efforts may be focused, is considering only fundamental, that is {\it truly} UV complete, QCD axion models. This is because fundamental field theories, such as asymptotically free~\cite{Gross} or asymptotically safe~\cite{WeinbergAS} models, have typically the ability to predict the value of some observables (see e.g. Ref.~\cite{Giudice:2014tma}).

Recently, the first fully calculable and realistic QCD axion model of this sort has been constructed~\cite{Salvio:2020prd}. It features a new non-Abelian interaction and new quarks and scalars. This model is valid up to infinite energy because is totally asymptotically free (TAF): all couplings can flow to zero in the infinite energy limit. The requirement of total asymptotic freedom is a sufficient but non-necessary feature of a fundamental theory, as asymptotic safety can be an alternative (see~\cite{Mann:2017wzh,Abel:2017rwl,Pelaggi:2017abg,Ghoshal:2017egr} for phenomenological applications). However, having all couplings approach zero in the UV allows us to trust perturbation theory, at least for sufficiently high energies, and thus obtain a calculable model.

This asymptotically free axion model, as is typically the case in fundamental field theories, predicts some low-energy observables: this is because some couplings (specifically the Yukawa and quartic couplings) are compatible with the TAF requirement only if they acquire  some specific isolated values at low energy. In other worlds, these couplings are IR attractive.   

 The requirement of having a classically scale-invariant axion sector, with no dimensionful parameters in the classical Lagrangian, further reduce  the number of independent parameters. Indeed, the   mass scales are then obtained quantum mechanically rather than with additional mass parameters in the classical Lagrangian. 
  For this reason we investigate here a region of the parameter space of the TAF axion model\cite{Salvio:2020prd} where the PQ symmetry is broken quantum mechanically through the Coleman-Weinberg (CW) mechanism~\cite{Coleman:1973jx}. This mechanism allows us, among  other things, to generate scales via quantum corrections in the regime of validity of perturbation theory and, therefore, have a fully calculable setup.

The main purpose of this work  is to study the PQ phase transition and the characteristics of the possible associated GWs  in this highly predictive framework. As pointed out in Ref.~\cite{Witten:1980ez} (see~\cite{DelleRose:2019pgi,vonHarling:2019gme} for recent applications to QCD axion models) the phase transitions associated with the CW symmetry breaking are typically  of first order and can, therefore, lead to observable GWs\footnote{See e.g. Refs.~\cite{Jaeckel:2016jlh} for previous studies of GWs and phase transitions in effective models with CW symmetry breaking.}. This, together  with the high predictivity of the TAF axion model mentioned above,  can lead  to testable implications at GW detectors. Furthermore, a typical feature of CW phase transitions is the presence of a phase of strong supercooling, meaning that the temperature below which the phase transition is effective (the nucleation temperature) turns out to be much smaller than the critical temperature \cite{Witten:1980ez} and, for the PQ symmetry breaking, much below $f_a$~\cite{DelleRose:2019pgi,vonHarling:2019gme}.

 Since the TAF requirement regards the extrapolation of the theory at arbitrarily high energies, some comments about the behavior of gravity in the UV are now in order. Here we assume that the gravitational interactions are  softened (compared to their behavior in Einstein's theory)  above and only above a certain energy scale $\Lambda_G$~\cite{Giudice:2014tma}. While at large  lengths all successes of Einstein's theory are reproduced, gravity is assumed to be so weak from the UV down to the PQ scale that its impact on the renormalization group equations (RGEs) can be neglected. This is possible because $\Lambda_G$ can be much below the Planck scale $\bp$, where quantum gravity effects in Einstein gravity would become sizeable. Since a phase of strong supercooling occurs in our classically scale-invariant TAF axion model, the nucleation temperature is much smaller than $f_a$ and the relevant values of the fields and their derivatives  are also much smaller than  $f_a$.
 %checked numerically
  This implies that gravity, as far as the production of gravitational waves is concerned, is well-described by Einstein's theory for all values of $\Lambda_G$ satisfying $f_a \lesssim\Lambda_G\ll\bp$.

This softened-gravity scenario may be realized, for example, in UV modifications of gravity featuring quadratic curvature terms in the action~\cite{Salvio:2014soa,Salvio2} or in non-local extensions of general relativity~\cite{Frolov:2015bta}. Interestingly, the former case also admits a classically scale-invariant formulation, called Agravity~\cite{Salvio:2014soa}, in which the Planck and the Fermi scales as well as the  cosmological constant are generated quantum mechanically via a gravitational version of the CW mechanism.

Given that gravity is softened for all energies above $\Lambda_G$ and Einstein's gravity is already very weak much below $\bp$, in this scenario all gravitational contributions to the effective action can be well described by perturbation theory.  This means that the non-perturbative Planckian corrections expected to spoil all global symmetries in Einstein's theory are actually negligible in the context of softened gravity. In particular the non-perturbative effects (described by Euclidean wormholes) that violate the PQ symmetry~\cite{Abbott:1989jw} and lead to the so-called PQ quality problem~\cite{Georgi:1981pu} in Einstein's gravity  can be neglected here. Therefore, softened gravity provides automatically a solution of the PQ quality problem.

 The paper is organized as follows. In Sec.~\ref{model} we introduce the TAF axion model without a fundamental PQ scale and review results regarding the RG flow in~\cite{Salvio:2020prd} that are most relevant for our purposes. 
In Sec.~\ref{Dynamical generation} it is shown  that  $f_a$  is generated quantum mechanically through the CW mechanism  and the mass spectrum is also discussed.  In the same section we also show that the axion sector only features one independent mass scale, i.e. $f_a$, and one independent dimensionless parameter (the QCD gauge coupling evaluated at $f_a$). The PQ phase transition is then studied in Sec.~\ref{PQ-PT}, which gives details about the finite-temperature effective potential, the bounce solutions and their actions, the nucleation temperature and the reheating after the supercooling era. In Sec.~\ref{GWs} the predictions of the classically scale-invariant TAF axion model regarding the GW spectrum are worked  out. A description of the relevant GW detectors is then provided in Sec.~\ref{GW_det}, where we also compare the sensitivities of the  GW experiments with the predictions of our fundamental model. Finally, in Sec.~\ref{Conclusions} we give our conclusions.

\section{A fundamental axion model without a  fundamental PQ scale}\label{model}

Here we consider a dimensionless version of the TAF axion sector of~\cite{Salvio:2020prd}: we take the limit in which the dimensionful parameter in the microscopic Lagrangian  go to zero. The axion sector is  invariant under an SU(2) group (henceforth SU(2)$_a$). Then the full gauge group contains the  factor SU(3)$_c\times$SU(2)$_a$, where SU(3)$_c$ is the ordinary QCD group.  
%In addition to  SU(3)$_c\times$SU(2)$_a$ 
The gauge group should also include extra factors to account for a TAF extension of the  SM   (see e.g.~\cite{Giudice:2014tma,Holdom:2014hla,Pelaggi:2015kna}). We call such an extension the ``SM sector", which of course  has to be present, in addition to  the axion sector we describe here in order for the complete model to be fully viable.
We will not commit ourselves to a specific TAF SM extension in this work, but we note that in general the SM and axion sectors interact via SU(3)$_c$ gauge interactions.
 
The model features two extra Weyl fermions $q$ and $\bar q$ in the fundamental and antifundamental of  SU(3)$_c\times$SU(2)$_a$,  with  the same PQ charge: $\{q,\bar q\}\rightarrow e^{i\gamma/2}\{q,\bar q\}$, where $\gamma$ is a constant. The PQ charges of all particles in the SM sector  vanish for simplicity, like in axion models of the KSVZ~\cite{Kim:1979if} type.  We introduce a scalar field $A$, which spontaneously breaks the PQ symmetry (denoted here U(1)$_{\rm PQ}$) and gives mass to the extra quarks (as required by the experiments).  Therefore, $A$ is complex and have Yukawa interactions with $q$ and $\bar q$,
\be  \mathscr{L}_y = - y \bar qA q +{\rm H.c.}\, . \label{Yukawa} \ee
The PQ symmetry implies that $A$ transforms  under U(1)$_{\rm PQ}$  as $A\rightarrow e^{-i\gamma}A$. Gauge invariance, instead, tells us that $A$ is invariant under SU(3)$_c$ and belongs to the adjoint of SU(2)$_a$. The scalar $A$, being complex, can be written as $A= A_R+i A_I$ where  $A_R$ and $A_I$ are Hermitian adjoint representations. Further Yukawa interactions besides~(\ref{Yukawa}) and those present  in the SM sector are forbidden by the gauge symmetries and U(1)$_{\rm PQ}$.

The potential of $A$ is given by
\be V_A=\lambda_1 {\rm Tr}^2(A^\dagger A)  +
 \lambda_2|{\rm Tr}(A A)|^2.
\label{UA}\ee
Note that $V_A$, just like any other term in the Lagrangian, only contains dimensionless coefficients. The possible mass term $m^2$Tr$(A^\dagger A)$ in the potential in~\cite{Salvio:2020prd} has been erased or, more generally speaking, it will be assumed that $m$ is much smaller than the effective mass generated by the CW mechanism (we will  discuss how this mechanism works in the present model in Sec.~\ref{Dynamical generation}).
The necessary {\it and} sufficient conditions for vacuum stability at high-field values (henceforth ``high-field stability") are~\cite{Salvio:2020prd}
\be \lambda_1>0, \qquad \lambda_1+ \lambda_2>0.\label{VacS12}\ee
Here we neglect the couplings with the scalars of the SM sector; setting those couplings exactly to zero is consistent at one-loop level. At higher-loop level those couplings can be generated, but they remain small and will have negligible effects on our  results.

%A VEV $\langle A\rangle$ breaks ${\rm U(1)}_{\rm PQ}$ and the gauge group SU($N$)$_a$ down to ${\rm U(1)}_a\times ...$, where ${\rm U(1)}_a$ is the Abelian gauge group generated by the generator along which $\langle A\rangle$ lies. This leads to extra massless gauge bosons besides the photon.
%

%\section{The RG flow}

Let us now review the beta-functions of this model~\cite{Salvio:2020prd}. The renormalization group equation (RGE) of the gauge coupling $g$  of a generic gauge group is
\be \frac{dg^2}{dt} = -bg^4, \label{betag}\ee 
where $t\equiv \ln(\mu^2/\mu_0^2)/(4\pi)^2$, the quantity $\mu_0$ is an arbitrary reference energy and $\mu$ is the usual RG scale. The  solution to Eq.~(\ref{betag}) is UV attractive
for any $g_0\equiv g(0)$ and AF requires $b>0$. The value $g=0$ is a trivial fixed point of Eq.~(\ref{betag}) and so we take $g>0$ without loss of generality.  For SU(2)$_a$ and SU(3)$_c$
the constant $b$ for the corresponding gauge couplings $g_a$ and $g_s$ reads, respectively,
\be b_a=\frac{14}{3}, \qquad b_s=\frac{29}{3}-\Delta,  \label{gsRGE}\ee
where $\Delta$ is the positive extra contribution due to the fermions and scalars in the SM sector. In the numerical calculation we will use for definiteness the reference value $\Delta=28/3$, which is compatible with known TAF SM extensions~\cite{Pelaggi:2015kna,Salvio:2020prd}.
The RGE of $y$ is instead
\be\frac{dy^2}{dt} = y^2\left(\frac{9 y^2}{2}-8 g_s^2-\frac{9 g_a^2}{2} \right).\label{yRGE}\ee
Like for the gauge couplings, the beta-function vanishes at zero coupling, 
%value $y=0$ is a trivial fixed point of Eq.~(\ref{yRGE}) and 
so we take $y>0$ without loss of generality.
This equation admits a closed-form solution for any $b_a$ and $b_s$~\cite{Salvio:2020prd}. Finally, the RGEs of $\lambda_1$ and $\lambda_2$ are $\frac{d\lambda_1}{dt}  = \beta_1$, and $\frac{d\lambda_2}{dt} =\beta_2$,
where~\cite{Salvio:2020prd} 
\be \beta_1(g_a, y, \lambda) =\frac92 g_a^4+\lambda_1 \left(8 \lambda_2+6 y^2-12 g_a^2\right)+14 \lambda_1^2+8 \lambda_2^2-3 y^4\label{lambda1RGE}\ee
and 
%\be \beta_2(g_a, y, \lambda_1,\lambda_2)  =\frac12\left[3 g_a^4+\lambda_2 \left(24 \lambda_1+12 y^2-24 g_a^2\right)+12 \lambda_2^2 +3 y^4\right]. 
% \ee
 \be \beta_2(g_a, y, \lambda)  =\frac32 g_a^4+\lambda_2 \left(12 \lambda_1+6 y^2-12 g_a^2\right)+6 \lambda_2^2 +\frac32 y^4.  \label{lambda2RGE}
 \ee
% Unlike the gauge and Yukawa couplings, the vanishing values $\lambda_1=0$ and/or $\lambda_2=0$ are not fixed points of the RGEs in~(\ref{lambda1RGE}) and~(\ref{lambda2RGE}), unless the gauge and Yukawa couplings vanish at the same energy. 
 Note that $\lambda_1=0$  and $\lambda_2=0$ generically are not zeros of $\beta_1$ and $\beta_2$  and, therefore, the quartic couplings, or some combinations thereof, could a priori change sign and pass through zero during the RG evolution.
 
%The RGEs of  the $\lambda_i$ are too complicated for us to determine analytically the general solution at any $t$, but they can be solved numerically.
 In~\cite{Salvio:2020prd} the system of  equations~(\ref{betag})-%,~(\ref{yRGE})~(\ref{lambda1RGE}) and~
 (\ref{lambda2RGE}) was solved and it was found that for any initial condition for the gauge couplings there is one and only one TAF solution satisfying the stability conditions in~(\ref{VacS12}) at high-field values.  For such solution both $y$ and $\lambda_i$   are   IR attractive and are, therefore, predicted at low energies~\cite{Salvio:2020prd}. We will pick up this TAF solution from now on  as the high-field stability conditions in~(\ref{VacS12})  are necessary to have a viable setup.

\section{Quantum generation of $f_a$}\label{Dynamical generation}

The RGEs dictate that the couplings run with energy.  The conditions in~(\ref{VacS12}) are necessary at high energy for high-field stability, but they can be violated in the  IR. In the present model we find that, while the first condition is always preserved, the second one is violated at small energy. This leads to spontaneous symmetry breaking of U(1)$_{\rm PQ}$ through the CW mechanism~\cite{Coleman:1973jx}, because at the energy scale $\mu_{\rm PQ}$ where $\lambda\equiv\lambda_1+\lambda_2=0$ the effective potential develops a flat direction (corresponding to $A=A^\dagger$, such that the two terms in the potential~(\ref{UA}) become equal: ${\rm Tr}(A^\dagger A)={\rm Tr}(A A)$).  We interpret   $\mu_{\rm PQ}$ as the PQ scale. On the flat direction $A=A^\dagger$ the three components $A_k$ of $A$ along the Pauli matrices, $A=A_k\sigma^k/2$, can always be transformed through an element of SU(2)$_a$ in a way that only one of these components is not vanishing and positive. We call this non-zero component  $\phi$. For example, a possible choice for $\phi$ is $\phi=|$Re$(A_1)|$. In~\cite{Salvio:2020prd} the case in which the explicit mass $m$ is larger than $\mu_{\rm PQ}$ was considered. Here we focus instead on the opposite case $m\ll \mu_{\rm PQ}$, so the PQ symmetry breaking is entirely driven by the CW mechanism.

A non-vanishing value of $\phi$ breaks SU(2)$_a$ down to a residual Abelian group U(1)$_a$ leading to a massless spin-1 particle (a dark photon), two spin-1 particles with equal mass 
\be M_V(\phi)=g_a \phi\label{MV}\ee 
(which can be described by one complex vector field), two degenerate Dirac fermions with mass 
\be M_Q(\phi) = y\phi/2, \label{MQ}\ee 
two scalars with squared mass 
\be M_S^2(\phi)=(\lambda_1-\lambda_2)\phi^2 \label{MS}\ee and two massless scalars (one is the axion, which as usual acquires a mass through quantum correction, and the other one corresponds to the flat direction). Note that $M_S^2\geq0$ when $\lambda_1\geq\lambda_2$, which turns out to be satisfied at all scales, from the TAF  requirement~\cite{Salvio:2020prd}.

At one-loop the quantum potential at zero temperature along the flat direction is given by
\be V_{\rm CW}(\phi) = V_0(\phi) + V_1(\phi),\ee
where $V_0$ is the tree-level potential, where  $\lambda$ is evaluated at the renormalization scale $\mu=\mu_0\exp(8\pi^2 t)$,
\be V_0(\phi) = \frac{\lambda(t)}{4} \phi^4, \ee
and $V_1$ is the quantum one-loop correction
\be V_1(\phi)=\sum_b \frac{n_bM_b(\phi)^4}{4(4\pi)^2}  \left(\ln\left(\frac{ M^2_b(\phi)}{\mu^2}\right) - a_b\right)-\sum_f \frac{n_f M_f(\phi)^4}{4(4\pi)^2}  \left(\ln\left(\frac{ M^2_f(\phi)}{\mu^2}\right) - a_f\right).\ee
In this expression the sum over $b$ runs over all bosons (with number of degrees of freedom $n_b$), that over $f$ runs over all fermions (with number of degrees of freedom $n_f$), $M_{b,f}(\phi)$ are the corresponding background-dependent masses (which, for our model, are given in Eqs.~(\ref{MV}),~(\ref{MQ}) and~(\ref{MS})) and $a_b$ and $a_f$ are renormalization-scheme dependent quantities. It is understood that the coupling constants in $V_1$ are evaluated at the same renormalization scale, $\mu$. This part of the potential can be computed explicitly by using the background-dependent masses given above. Setting $\mu=\mu_{\rm PQ}$,  where $\lambda$ vanishes, leads to
\be V_{\rm CW}(\phi) = \frac{\bar\beta}{4} \left(\ln\left(\frac{\phi}{f_a}\right)-\frac14\right)\phi^4, \label{CWpot}\ee 
where $\bar\beta$ is the beta-function of $\lambda$ evaluated at $\mu_{\rm PQ}$, namely
\be \bar\beta \equiv \left[\mu\frac{d\lambda}{d\mu}\right]_{\mu=\mu_{\rm PQ}}, 
\ee
and the scale $f_a$ has been introduced in a way that the CW potential, Eq.~(\ref{CWpot}), has its stationary point at $\phi=f_a$. We conventionally choose a renormalization scheme such that $f_a = \exp(-1/4) \mu_{\rm PQ}$. When $\bar\beta>0$ the stationary point at $\phi=f_a$ corresponds to a minimum. We have numerically verified the positivity of $\bar\beta$ in our model. Then U(1)$_{\rm PQ}$ is spontaneously broken and $\phi$ acquires the VEV $\langle \phi\rangle = f_a$ and a mass squared $M_\phi=\sqrt{\bar\beta}f_a$. 
The energy scale $f_a$ is, therefore, identified with the PQ symmetry breaking scale. The remaining mass spectrum besides $M_\phi$ corresponds to the axion, the dark photon and the tree-level masses  obtained   by setting $\phi=f_a$ in Eqs.~(\ref{MV}),~(\ref{MQ}) and~(\ref{MS}). Since $f_a$  above the scale of $10^8$~GeV and the mass of the extra-quarks  $M_Q(f_a)$ is around $f_a$, the dark photon satisfies all present bounds, including those of cosmological nature~\cite{Salvio:2020prd}.

\begin{figure}[t]
\begin{center}
\includegraphics[scale=0.55]{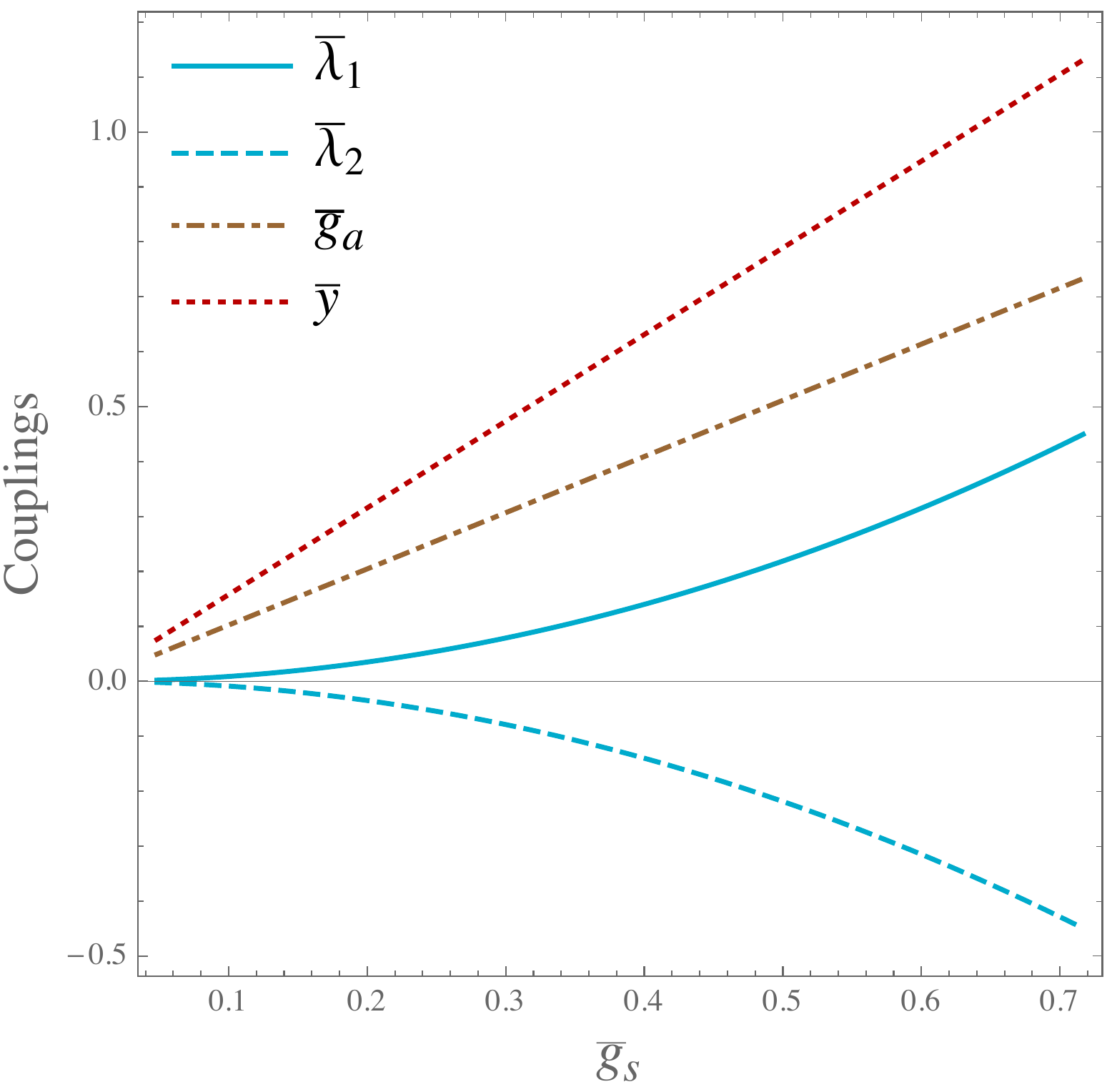} 
\end{center}
	\caption{\em  The couplings of the model as functions of the QCD gauge coupling at the PQ scale.}
\label{couplings}
\end{figure}

 Note that the arbitrariness of $\mu_0$ tells us that $\mu_{\rm PQ}$ (and thus $f_a$) is a free parameter. The TAF axion sector features only another free parameter, which can be taken to be $\bar g_s\equiv g_s(t_{\rm PQ})$ (in this paper a bar   indicates that a generic coupling is evaluated at 
$t=t_{\rm PQ}\equiv \ln(\mu_{\rm PQ}^2/\mu_0^2)/(4\pi)^2$). Indeed, 
%as long as $g_s$ and $g_a$ are non vanishing and remain perturbative, we can always shift $t$ (rescaling $\mu$)  in a way to set $g_{a0}$ equal to any positive number and the couplings $y$, $\lambda_1$ and $\lambda_2$ are predicted at low  energy~\cite{Salvio:2020prd}. 
%Another 
once the gauge couplings are chosen at $\mu=\mu_{\rm PQ}$ the other couplings $\bar y$, $\bar \lambda_1$ and $\bar \lambda_2$ are predicted~\cite{Salvio:2020prd} and one must consider a particular IR value of one of the gauge  couplings, say $\bar g_a$, to enforce $\bar\lambda_1+\bar \lambda_2 =0$, namely to have CW symmetry breaking.
% $\bar g_a$ is not another independent parameter because, once $\bar g_s$ is chosen, one must consider a particular value of $\bar g_a$ to enforce $\bar\lambda_1+\bar \lambda_2 =0$, while the other two couplings $\bar y$ and $\bar \lambda_1$ are predicted~\cite{Salvio:2020prd}.
  In Fig.~\ref{couplings} we give  the couplings $\bar g_a$, $\bar  y$ and $\bar\lambda_1 = -\bar\lambda_2$ as functions of $\bar g_s$ to show that all dimensionless quantities in the axion  sector are fixed once $\bar g_s$ is chosen. As a result, when the PQ symmetry is broken \`a la CW one also obtains a prediction for the mass of the extra complex vector field (in addition to the predictions  of the extra scalar and fermion masses% when $m$ is larger than $\mu_{\rm PQ}$
~\cite{Salvio:2020prd}) once $f_a$ and $\bar g_s$ are chosen. Therefore, we explicitly see that the CW mechanism to break U(1)$_{\rm PQ}$ leads to a more predictive framework than the symmetry breaking mechanism based on the explicit mass term $m^2$Tr$(A^\dagger A)$.

%read up to  here

\section{Peccei-Quinn phase transition}\label{PQ-PT}
In order to investigate the nature of the Peccei-Quinn phase transition we take into account thermal corrections as well as  quantum corrections. We consider the one-loop effective potential
\be V_{\rm eff}(\phi, T) \equiv V_{\rm CW}(\phi) + V_{T}(\phi)+\Lambda_0,  \ee
where the thermal correction $V_{T}$ to the effective potential is given by \cite{Dolan:1973qd} (see also \cite{Quiros:1994dr})
\be V_T(\phi) =  \frac{T^4}{2\pi^2} \left(\sum_b n_b J_B(M^2_b(\phi)/T^2)-\sum_f n_f J_F(M^2_f(\phi)/T^2)\right),  \ee
with  
\be J_{B,F}(x)\equiv \int_0^\infty dq \,  q^2\ln \left[1\mp\exp\left(-\sqrt{q^2+ x}\right)\right], \ee
and we have included  in	 $V_{\rm eff}(\phi, T)$ a constant term $\Lambda_0$ to account for the observed value of the cosmological constant when $\phi$ is at the minimum. It is understood that the coupling constants in $V_T$ are evaluated at the same renormalization scale, $\mu$, used in $V_{\rm CW}$.
 
 Since the background-dependent squared masses are all non-negative $V_{\rm eff}$ has a vanishing imaginary part. This is due to the fact that $f_a$ is generated  quantum mechanically rather than through an explicit tachyonic scalar mass, which would unavoidably lead to a concave tree-level potential and thus to a complex effective potential for some field values. Therefore, the CW  symmetry breaking supports the validity of perturbation theory: indeed, a non-negligible imaginary part (absent in  the CW case)  generically signals the breaking of the perturbative expansion. Further comments regarding the approximation  used will  be given below in  this section.
%We neglect the derivative corrections to the effective action as they are  small in our case (as we will check below).

\begin{figure}[t]
\begin{center}
\includegraphics[scale=0.65]{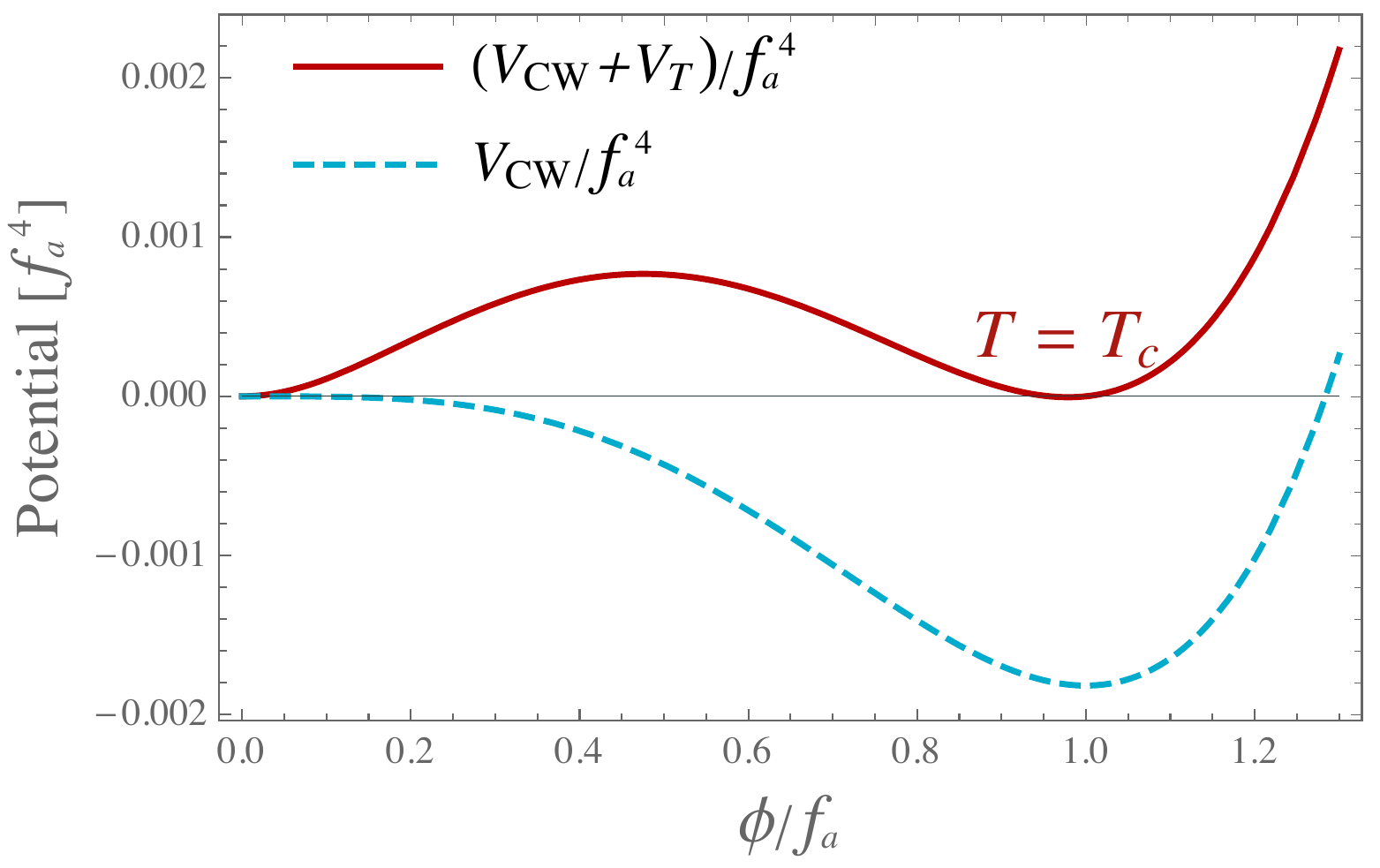} \hspace{0.8cm} \includegraphics[scale=0.59]{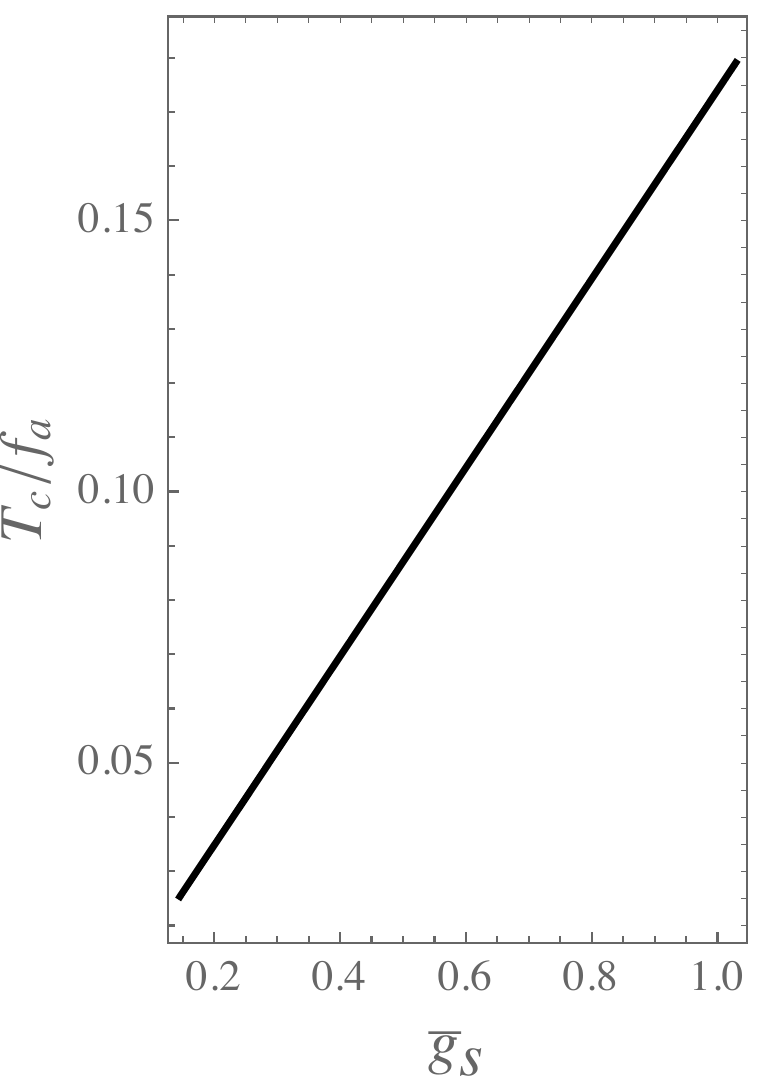} 
\end{center}
	\caption{\em  {\bf Left plot:} The effective potential setting the QCD gauge coupling at the PQ scale to $\bar g_s \approx 0.91$ and adding a constant such that it vanishes at $\phi=0$. {\bf Right plot:} the critical temperature $T_c$ divided by $f_a$ as a function of $\bar g_s$.}
\label{PT}
\end{figure}

In Fig.~\ref{PT} (left plot) we show  $V_{\rm eff}$ as a function of $\phi$  for two values of the temperature: the critical temperature $T_c$ and $T=0$. That figure shows that the transition is of first order. Although we use a fixed value of $\bar g_s$ in that figure, other choices of this parameter lead to  the same qualitative situation.  In the right plot of Fig.~\ref{PT} we give the dimensionless quantity $T_c/f_a$ as a function of the only dimensionless parameter of the axion sector, $\bar g_s$.

The absolute minimum of the effective potential is at $\langle \phi\rangle=0$ for $T>T_c$, while, for $T<T_c$, is at a non-vanishing temperature-dependent value. In the latter case the decay rate per unit volume $\Gamma$ of the false vacuum $\phi=0$ into the true vacuum $\phi=\langle\phi\rangle\neq 0$ can be computed with the formalism of~\cite{Coleman:1977py,Callan:1977pt,Linde:1980tt,Linde:1981zj}:
\be  \Gamma\approx \max\left(T^4\left(\frac{S_3}{2\pi T}\right)^{3/2}\exp(-S_3/T)\,, \,\, \frac{1}{R_4^4}\left(\frac{S_4}{2\pi}\right)^{2}\exp(-S_4)\right). \label{GammaOption}\ee
Here $S_{d}$ is the action 
\be S_d= \frac{2\pi^{d/2}}{\Gamma(d/2)} \int_0^\infty dr \, r^{d-1}\left(\frac12 \phi'^2+V_{\rm eff}(\phi,T)\right) \label{Sd1}\ee 
evaluated at the O($d$) bounce  defined as the solution of the  differential problem 
\be \phi''+\frac{d-1}{r}\phi'= \frac{dV_{\rm eff}}{d\phi}, \qquad \phi'(0)=0, \quad \lim_{r\to \infty}\phi(r) = 0,\ee
where a prime denotes a derivative with respect to $r$. 
Also, $R_4$ is the size of the $O(4)$ bounce. Note that $S_d$ evaluated at the O($d$) bounce can be simplified through the  scaling arguments of~\cite{Coleman:1977th} to obtain 
\be S_d =  \frac{4\pi^{d/2}/\Gamma(d/2)}{(2-d)} \int_0^\infty dr \, r^{d-1} V_{\rm eff}(\phi,T). \ee
Using the expression above instead of the one in~(\ref{Sd1}) makes numerical calculations easier  because the derivative of the bounce does not appear in  the action.

\begin{figure}[t]
\begin{center}
\includegraphics[scale=0.5]{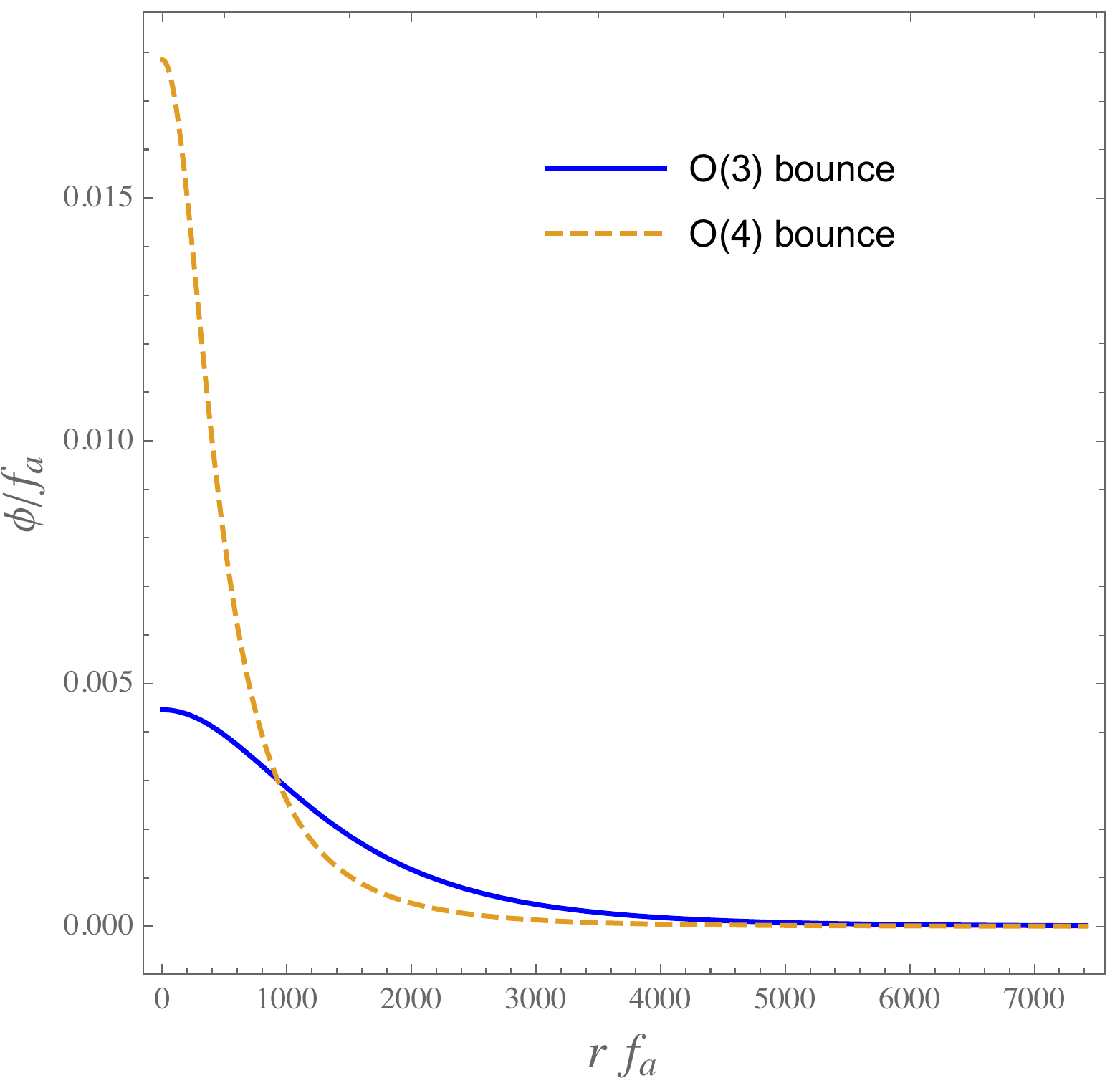} 
\end{center}
	\caption{\em  The O(d)-symmetric bounces for $\bar g_s = 0.91$  and $T \approx 4 \tt 10^{-3} \, T_c\approx 6.3\tt 10^{-4}f_a$.}
\label{Bounce34}
\end{figure}

We numerically checked that $S_3/T < S_4$, so $\Gamma$ is dominated by the O(3) bounce. Therefore, the phase transition is essentially  due to thermal effects rather than quantum effects. As an example, in Fig.~\ref{Bounce34} we give a plot of the O(3) and O(4) bounces for representative values of the QCD gauge coupling and the  temperature below $T_c$.

Some words on the approximation used are now in order. We note that the correction to the two-derivative term in the one-loop effective action is  small as long as the temperature is small compared to the field values~\cite{Moss:1985ve} characterising  the bounce solution. We have checked that $T$ is  small compared to (few \% of) the relevant field values for  all  numerical  calculations performed in this  work. This also implies that we are far from the high-temperature regime for which the perturbative expansion is known to break down~\cite{Weinberg:1974hy}. Moreover, note that, generically,  the higher-derivative corrections to the one-loop effective action are suppressed when the laplacian applied to the solution of interest (in this case the bounce) is small compared to the background dependent masses times that solution: this follows from the structure of the equations of motion without higher derivatives. Since the largest couplings are of order one in our case, those higher-derivative corrections are small when $\frac{dV_{\rm eff}}{d\phi}$ is small compared to $\phi^3$ in the relevant range of $r$ (where $S_3/T$ gets its dominant contribution). We numerically checked that this condition is also satisfied (at the few \% level). Therefore, our one-loop approximation for the effective action is reliable.

Also the gravitational corrections to the false vacuum decay are amply negligible in our case. This is because the typical scales of the bounce and the temperature are always below $f_a$ (as illustrated in Figs.~\ref{PT} and~\ref{Bounce34}) and the gravitational corrections are, therefore, suppressed by factors at least as  small as $f_a^2/\bp^2$~\cite{Salvio:2016mvj}. The latter quantity is tiny because $f_a$ is several orders of magnitude below $\bp$ in order for the axion not to overproduce DM (see Ref.~\cite{DiLuzio:2020wdo} for a review).

\begin{figure}[t]
\begin{center}
\includegraphics[scale=0.5]{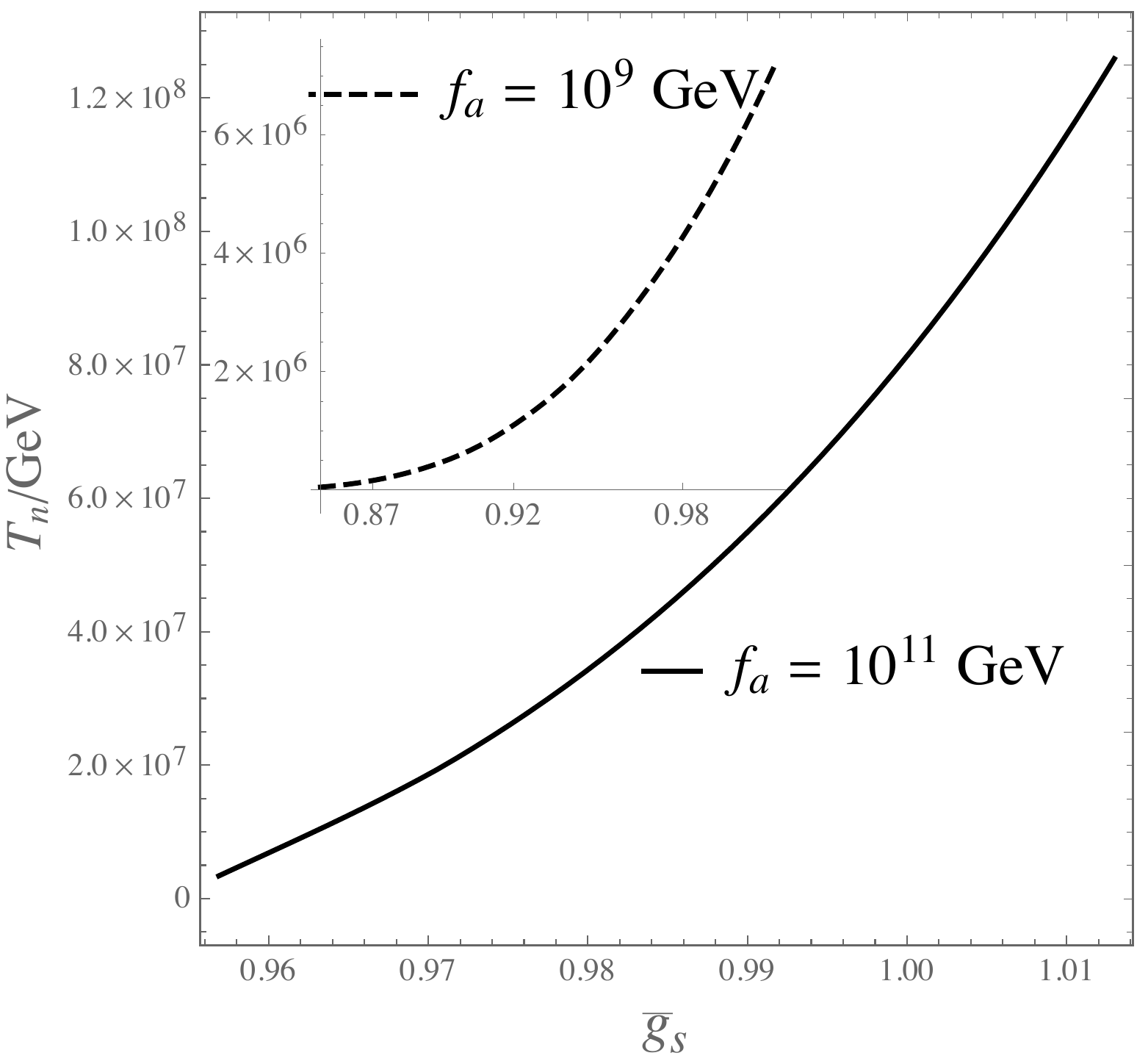} \hspace{0.9cm}
\includegraphics[scale=0.47]{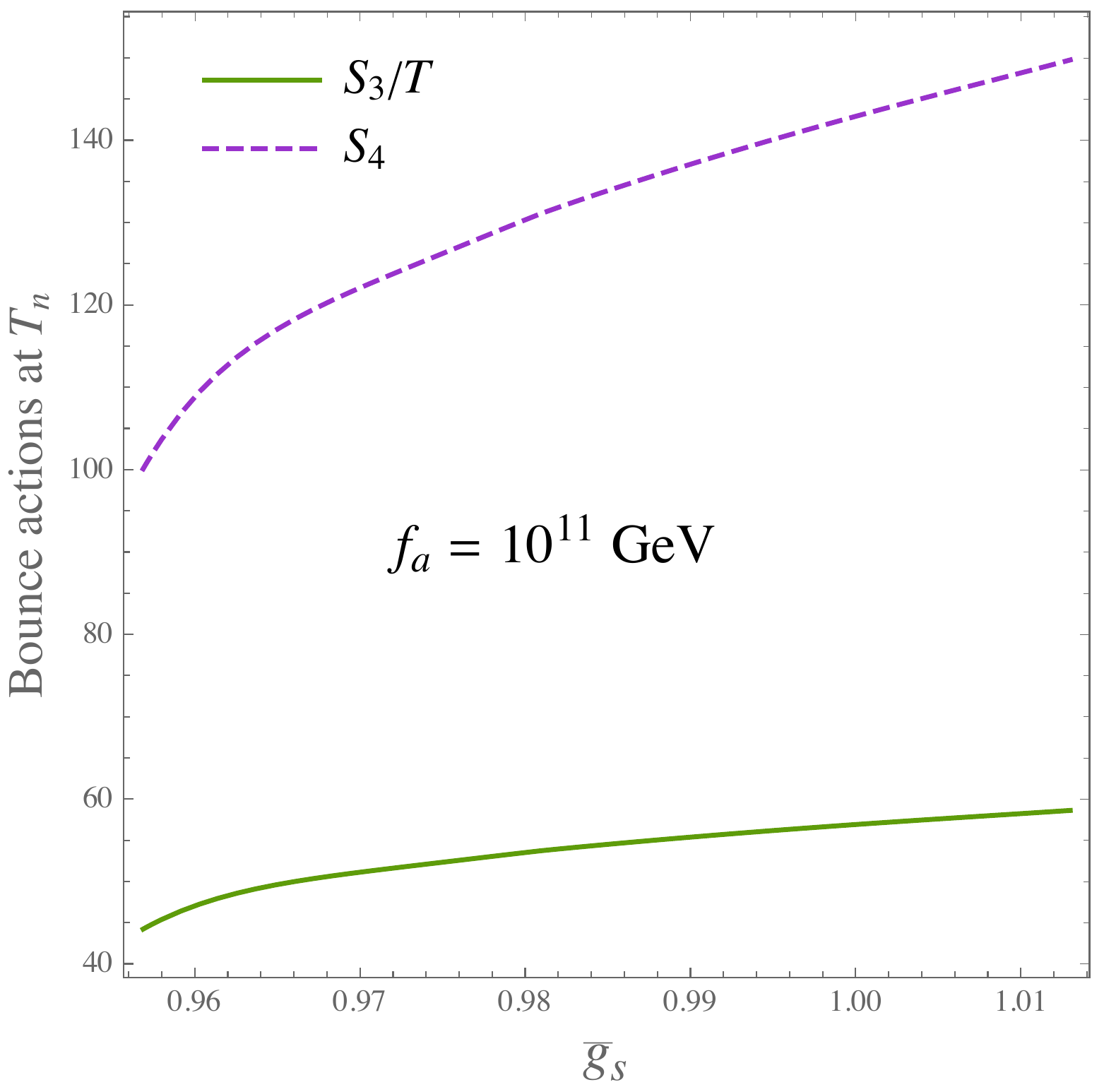} 
\end{center}
	\caption{\em  {\bf Left plot:} The nucleation temperature $T_n$ as a function of $\bar g_s$. {\bf Right plot:} the bounce actions $S_3/T$ and $S_4$ evaluated at $T_n$ as a function of $\bar g_s$.}
\label{Tnvsgs}
\end{figure}

Like for other models with CW symmetry breaking~\cite{DelleRose:2019pgi,vonHarling:2019gme}, we find that when $T$ goes below $T_c$ the scalar field $\phi$ is trapped in the false vacuum $\phi=0$ until  $T$ is much below $T_c$, in other words the universe features a phase of strong supercooling. 
Then the energy density  is dominated by the vacuum energy of $\phi$ and the universe grows exponentially  like during inflation but with Hubble rate $H_I = \sqrt{\bar\beta} f_a^2/(4\sqrt{3}\bp)$, where $\bp$ is the reduced Planck mass that is defined in terms of the Planck mass $M_P$ by $\bp \equiv M_P/\sqrt{8\pi}$. The bubbles created are   diluted by the expansion of the universe and they cannot collide until $T$ reaches the nucleation temperature $T_n$, which corresponds to the temperature when $\Gamma/H_I^4 \sim 1$ or, equivalently, using the fact that the decay is dominated by the O(3) bounce,
\be  \frac{S_3}{T_n}-\frac32 \ln \left(\frac{S_3/T_n}{2\pi}\right) = 4 \ln \left(\frac{T_n}{H_I}\right).\label{TnEq}\ee
In Fig.~\ref{Tnvsgs} (left plot) we give $T_n$ as a function of $\bar g_s$  for two physically interesting values of the PQ symmetry breaking scale: $f_a=10^{11}~$GeV (for which the axion can account for the whole DM) and $f_a=10^9~$GeV (for which the contribution of the axion to the DM abundance is negligible)\footnote{For $f_a=10^9~$GeV DM can be accounted for by other extra fields among those that are compatible with the TAF principle in the  SM  sector (see e.g.~\cite{Giudice:2014tma,Holdom:2014hla,Pelaggi:2015kna}).}. We find that the equation that determines  $T_n$ in~(\ref{TnEq}) does not always admit a solution:  there is a minimal value of the coupling $\bar g_s$ below which there is no solution (see the left plot of Fig.~\ref{Tnvsgs}). As clear from the left plot of Fig.~\ref{Tnvsgs}, when the coupling goes to this minimal value   $T_n$ becomes very small. By comparing the right plot of Fig.~\ref{PT} with the left plot of Fig.~\ref{Tnvsgs} one can see that supercooling generically takes place, namely $T_n\ll T_c$.  In Fig.~\ref{Tnvsgs} we also plot the bounce actions $S_3/T$ and $S_4$ evaluated at $T_n$ as functions of $\bar g_s$. That plot shows, as mentioned above,  that $S_4>S_3/T$ and so the phase transition is dominated by thermal effects. In the figure we consider, as an example, $f_a =10^{11}~$GeV, but the other values of $f_a$ lead to the same qualitative behavior. Moreover, an important requirement is that the phase transition completes. The detailed conditions for the true vacuum to fill the entire space were found in~\cite{Ellis:2018mja} and~\cite{Ellis:2020awk}. We have checked that this requirement is satisfied for all numerical calculations performed in this paper.
%For example, for $\{\bar g_s,f_a\} \approx  \{0.91,1.2 \tt 10^9\,{\rm GeV}\}$ we find $T_n \approx 7.4\tt 10^5$\,GeV $ \ll f_a$, while for $\{\bar g_s,f_a\} \approx  \{0.97, 10^{11}\,{\rm GeV}\}$ we obtain $T_n \approx 1.7 \tt 10^7$\,GeV $ \ll f_a$.

We also note that strong supercooling and the corresponding inflationary period efficiently dilute the density $n(T)$ of monopoles\footnote{See also Refs.~\cite{Sato:2018nqy} for a discussion on monopole production in the absence of supercooling.} due to the breaking SU(2)$_a \to$ U(1)$_a$. In a strong first-order phase transition monopoles may be created by bubble collisions and well-known estimates~\cite{Preskill:1979zi} lead to
\be \frac{n(T_n)}{T_n^3}  \gtrsim p \left(\frac{T_n}{C M_P}\right)^3,\label{Mtbound}\ee
where $p$ is the probability that the scalar field configuration is topologically non trivial, $C= 0.6/\sqrt{g_*(T_n)}$ and $g_*(T)$ is the effective number of relativistic species in thermal equilibrium at temperature $T$. Even for $p\approx 1$, setting $g_*(T_n)$ of order $10^2$ (a realistic setup given the existing TAF SM sectors) we find that the theoretical bound in~(\ref{Mtbound}) is amply compatible (unlike in grand unified theories without inflation) with the bound coming from the fact that the mass density of monopoles must not exceed the limit on the total mass density  imposed by the observed Hubble constant and deceleration parameter~\cite{Preskill:1979zi}. Indeed, the latter bound is around $n(T_0)/T_0^3 \lesssim 10~$eV$/M_m$, where $T_0$ is today's temperature, $M_m$ is the monopole mass and $M_m\sim 4\pi f_a/\bar g_a$, ~$n(T_0)/T_0^3  \lesssim  n(T_n)/T_n^3$ and the window $10^8~$GeV$~\lesssim f_a \lesssim 10^{12}~$GeV have been used.

The strength of the phase transition is measured, as usual, by the parameter $\alpha$ defined as the ratio between 
%the free-energy density associated with the transition 
\be \rho(T_n) \equiv \left[\frac{T}{4} \frac{d}{dT} \Delta V_{\rm eff}(\langle\phi\rangle,T)-\Delta V_{\rm eff}(\langle\phi\rangle,T)\right]_{T=T_n}, \ee
where $\Delta V_{\rm eff}(\langle\phi\rangle,T)\equiv V_{\rm eff}(\langle\phi\rangle,T)-V_{\rm eff}(0,T)$, and the energy density of the thermal plasma (see~\cite{Caprini:2019egz,Ellis:2019oqb} for more details). So
\be \alpha \equiv \frac{30 \rho(T_n)}{\pi^2 g_*(T_n)T_n^4}. \ee
 We find a very strong phase transition with $\alpha$ exceeding one by several orders of magnitude.

At the end of supercooling the universe should be reheated. This occurs in general  thanks to the unavoidable coupling between the axion sector and the SM sector due to gluons. The field $\phi$ couples at one loop to gluons through the extra quarks $q$ and $\bar q$ so one gets an effective interaction 
\be O_{\rm eff} \sim \frac{\bar y\bar g_s^2}{(4\pi)^2}\phi G_{\mu\nu}G^{\mu\nu}/M_Q,\ee where $G_{\mu\nu}$ is the gluon field strength. This leads to the following rate of the decay of $\phi$ into two gluons
\be \Gamma_{\phi\to gg} \sim \frac{\bar y^2\bar g_s^4 M_\phi^3}{(4\pi)^5M_Q^2}. \ee
%For example, for $\bar g_s = 0.91$ and $f_a\approx 1.2 \tt 10^9$\,GeV 
We find $\Gamma_{\phi\to gg}\gg H_I$ so the reheating is approximately instantaneous.

The reheating temperature due to this channel may be computed through
\be T_{\rm RH} = \left(\frac{45 \Gamma^2_{\phi\to gg}\bp^2}{4\pi^3 g_*(T_{\rm RH})}\right)^{1/4}.\ee
But  this formula  is only valid if the radiation energy density $\rho_R$ does not exceed the vacuum energy density $\rho_{\rm vac}$ due to $\phi$ (because $\rho_{\rm vac}$ represents the full energy budget of the system). If this condition is not satisfied we determine $T_{\rm RH}$ as the maximal temperature compatible with $\rho_R\leq \rho_{\rm vac}$,  leading to 
\be T_{\rm RH}^4 \approx \frac{15 \bar\beta f_a^4}{8\pi^2 g_*}.\label{TRHmax}\ee 
Our estimate of $T_{\rm RH}$ agrees to very good accuracy with previous determinations~\cite{DelleRose:2019pgi}. We find a very high $T_{\rm RH}$. For example, setting  $g_*\sim 10^2$  and $\{\bar g_s,f_a\} \approx  \{0.91,1.2 \tt 10^9\,{\rm GeV}\}$ we obtain $T_{\rm RH} \sim 10^8$\,GeV, while for $\{\bar g_s,f_a\} \approx  \{0.97, 10^{11}\,{\rm GeV}\}$ we obtain $T_{\rm RH} \sim 10^{10}$\,GeV. Note, nevertheless, that the maximal value of $T_{\rm RH}$  in~(\ref{TRHmax}) implies that $T_{\rm RH}$ is always below $f_a$ because $\bar\beta$ is a beta function and thus loop suppressed and $g_*$ is at least of order $10^2$.
%Note that the fact that supercooling is strong leads to the estimate $T_{\rm RH}^4 \approx 15 \bar\beta f_a^4/(8\pi^2 g_*)$ (as discussed in~\cite{DelleRose:2019pgi}), 
%$T_{\rm RH} \sim 480/(\pi^2 g_* \beta f_a^4)$, 
%which agrees to very good accuracy with our estimate based on the decay of $\phi$.

\begin{figure}[t]
\begin{center}
\includegraphics[scale=0.69]{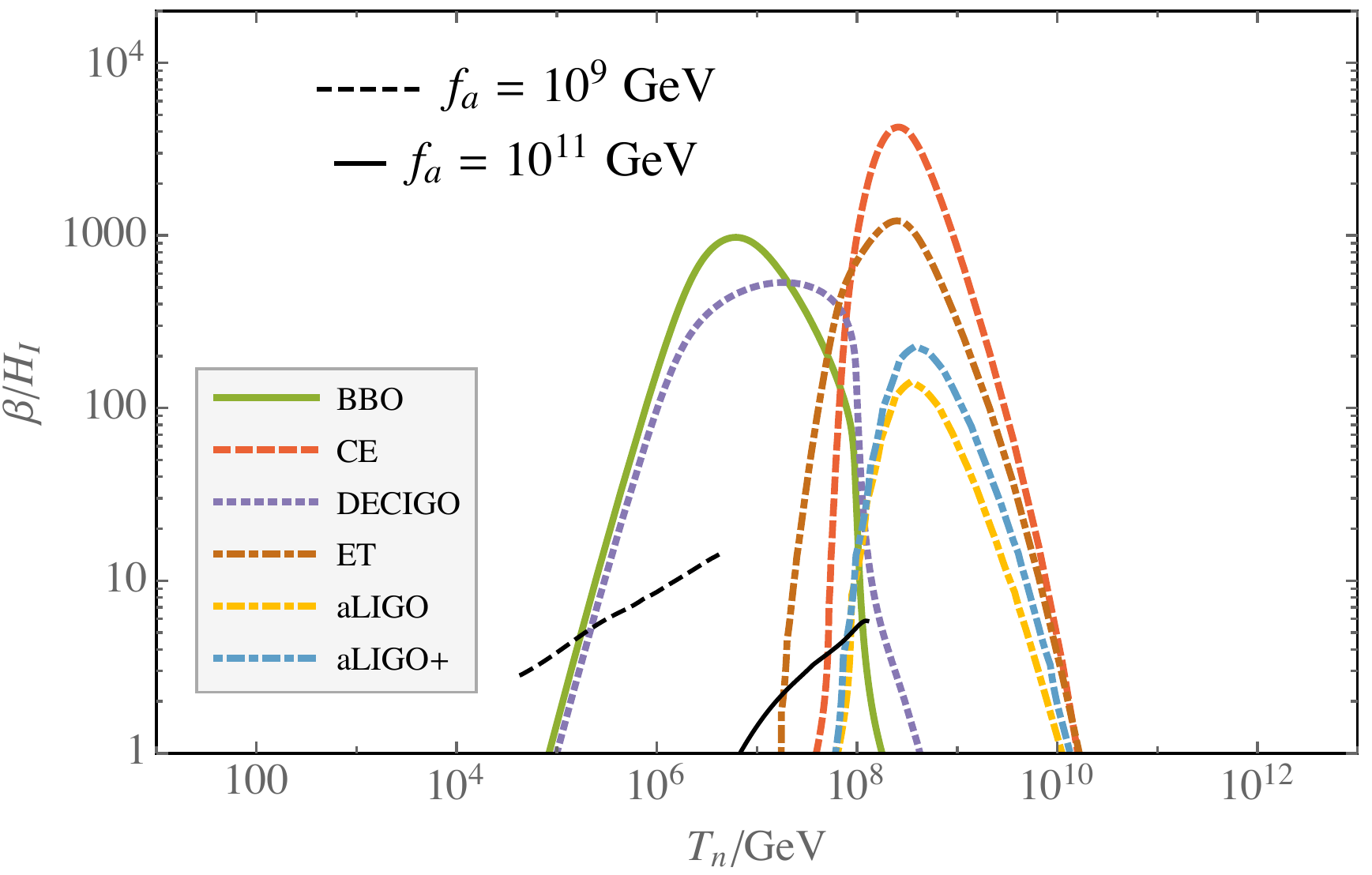}
%
%\vspace{0.6cm}
 %\includegraphics[scale=0.62]{bht_1} 
%\includegraphics[scale=0.5]{bht_3} 
\end{center}
	\caption{\em  The theoretical prediction  in the ($ T_n, \beta/H_I $) plane compared with the sensitivity curves of BBO, CE, DECIGO, ET and advanced LIGO (see Sec.~\ref{GW_det}).
	% (upper plot) and LISA (lower plot) 
	The theoretical curves are produced by varying $\bar g_s$ as in Fig.~\ref{Tnvsgs}. The parameter space enclosed within the experimental curves represents detectable signals.}
	\label{beta_H_T}
\end{figure}

 Finally, another important parameter   is the inverse of the duration of the phase transition which, following Ref.~\cite{vonHarling:2019gme}, we define through
 \be\beta \equiv \left[\frac1{\Gamma}\frac{d\Gamma}{dt}\right]_{T_n}.\ee This quantity, for fast reheating, can be computed with the formula~\cite{vonHarling:2019gme}
%see [von Harling - Pomarol - Pujolàs - Rompineve 1912.07587] 
\be \frac{\beta}{H_I} \approx \left[T\frac{d}{dT}(S_3/T)-4\right]_{T=T_n}, \ee 
where the  term $-4$ is due to the fact that $\Gamma\propto T^4$, because, as we checked, the phase transition is dominated by thermal effects and the extra temperature dependence of $\left(\frac{S_3}{2\pi T}\right)^{3/2}$ in~(\ref{GammaOption}) can be neglected because only logarithmic in the CW case.
For the reference values $\{\bar g_s,f_a\} \approx  \{0.91,1.2 \tt 10^9\,{\rm GeV}\}$ and $\{\bar g_s,f_a\} \approx  \{0.97, 10^{11}\,{\rm GeV}\}$ we obtain $\beta/H_I \approx 7$ and $\beta/H_I \approx 2$, respectively. The quantity $\beta/H_I$ for other values of $\{\bar g_s,f_a\}$ is given in Fig.~\ref{beta_H_T}. The relatively small values of $\beta/H_I$ that we obtain indicate a fairly long phase transition. In the same plot we also compare these theoretical values of $\beta/H_I$ with the experimental sensitivities (see Sec.~\ref{GW_det}).

We conclude this section by mentioning that the SM sector might also feature further strong first-order phase transitions. This is due to the fact that the SM gauge group must be extended in order to satisfy the TAF requirement and, therefore, additional symmetry breaking patterns are present. We note, however, that it is possible and natural to expect these further phase transitions at temperatures somewhat below the TeV or 10 TeV scales because the corresponding new physics can be at those energies (see e.g.~\cite{Pelaggi:2015kna}). The {\it typical} temperatures of the PQ phase transition is instead much bigger as shown in Fig.~\ref{Tnvsgs} (left plot).

\section{Gravitational waves}\label{GWs}

When the temperature drops below $T_n$ GWs are produced. The dominant source of GWs are bubble collisions that take place in the vacuum. This is  because in the era when $T$ reaches $T_n$ the energy density is dominated since a long time by the vacuum energy density associated with $\phi$, which leads to an exponential growth of the cosmological scale factor as we have seen. This inflationary behavior  as usual dilutes preexisting matter and radiation and, therefore, we neglect the GW production due to turbulence and sound waves  in the cosmic fluid\footnote{We explicitly checked that the inclusion of the efficiency factors of Ref.~\cite{Ellis:2019oqb} gives a subdominant correction in our case.}~\cite{Maggiore:2018sht}.
%, which are the main source of GWs because supercooling is very strong. 

From~\cite{Caprini:2015zlo} (which used, among other things, the results of~\cite{Huber:2008hg}  based on the envelope approximation) we find the following GW spectrum 
%as a function of the frequency $f$ 
due to vacuum bubble collisions (valid in the presence of supercooling and $\alpha\gg 1$)
\be \label{eq:gw_col} h^2 \Omega_{\rm GW}(f) \approx 1.29%7
\tt 10^{-6}\left(\frac{H(T_{\rm RH})}{\beta}\right)^2\left(\frac{100}{g_*(T_{\rm RH})}\right)^{1/3}\frac{3.8(f/f_{\rm peak})^{2.8}}{1+2.8(f/f_{\rm peak})^{3.8}},\ee 
where  $f_{\rm peak}$ is the red-shifted frequency peak today and is given by~\cite{Caprini:2015zlo} 
\be f_{\rm peak} \approx 3.79%83
 \tt 10^2\frac{\beta}{H(T_{\rm RH})}\frac{T_{\rm RH}}{10^{10}{\rm GeV}} \left( \frac{g_*(T_{\rm RH})}{100}\right)^{1/6} \, {\rm Hz} . \ee
%This frequency is the peak frequency today, which is red-shifted from the time $t_{\star}$ it was produced during the transition with a frequency
%$f_{\star}$
%so $f_{\rm peak}  = \frac{a_{\star}}{a_0} f_{\star}$, where $a_0$ and $a_{\star}$ are the scale factor today and at $t_{\star}$, respectively. 
Being the reheating almost instantaneous we can approximate the Hubble rate at $T_{\rm RH}$ with its value $H_I$. Some progress has been made to compute the GW spectrum beyond the envelope approximation~\cite{Jinno:2017fby,Konstandin:2017sat,Lewicki:2020jiv}, but Eq.~(\ref{eq:gw_col}) remains to date a reasonable and simple  approximation for the bubble collisions that take place in the vacuum~\cite{Konstandin:2017sat} (what we are mainly interested in).

\begin{figure}[t]
\begin{center}
\includegraphics[scale=0.55]{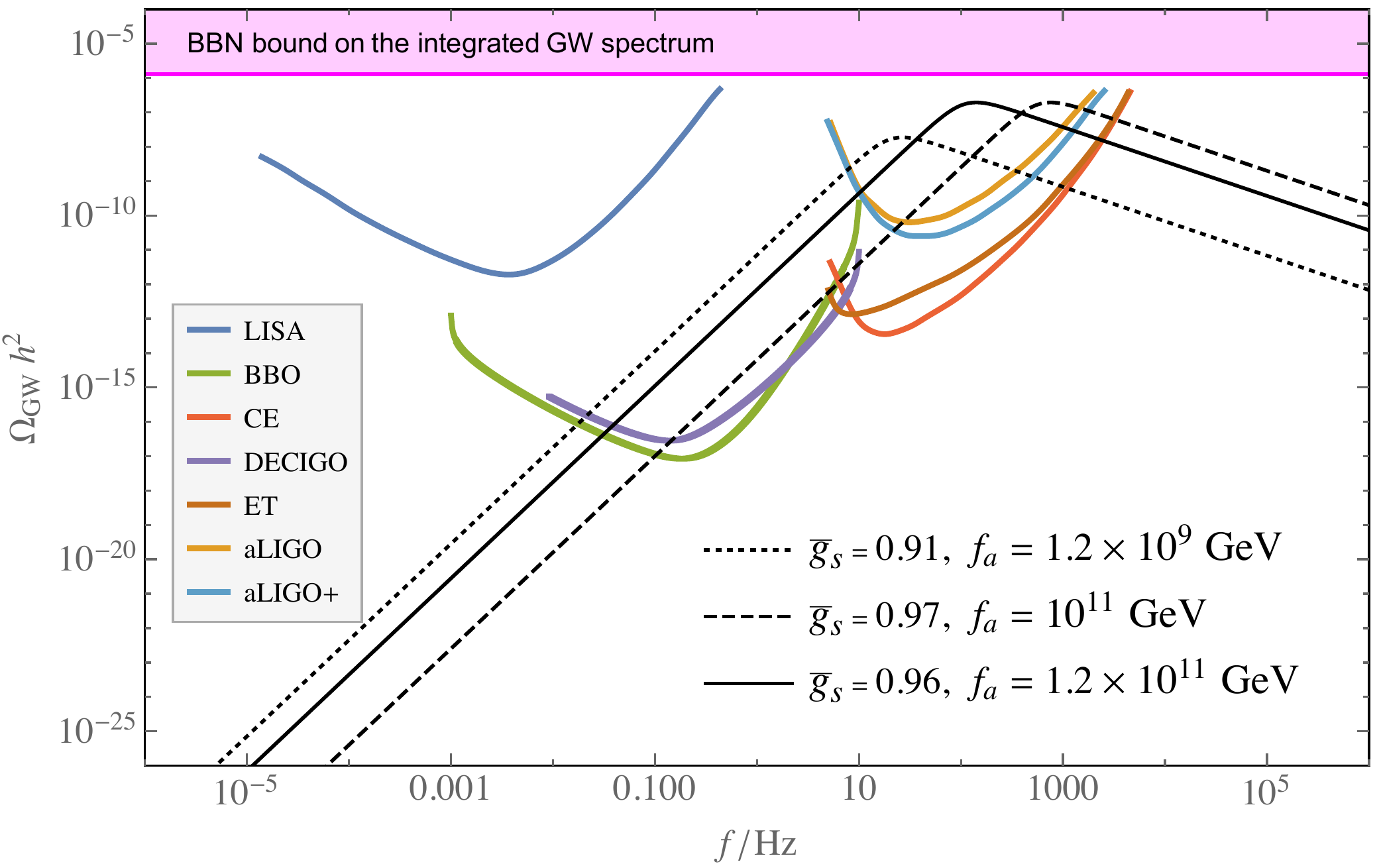}
\end{center}
	\caption{\em  The black lines give the GW amplitude corresponding to the values of the TAF model as mentioned in the figure.
	%$h^2\Omega_{\rm GW}(f)$ for $g_*=2\tt  10^2$ and $\{\bar g_s,f_a\} \approx  \{0.91,1.2\tt 10^9\,{\rm GeV}\}$.
	%and $\{\bar g_s,f_a\} \approx  \{0.97,10^{11}\,{\rm GeV}\}$. 
	The areas above the colored lines correspond to the projected sensitivities  for various GW observatories as detailed in section \ref{GW_det}.  The colored region represents the BBN bound  on the integrated GW spectrum in~(\ref{BBNbound}) for the reference upper bound $N_{\rm eff} =3.28$. 
		}
\label{Omega}
\end{figure}

In Fig.~\ref{Omega} we give $h^2\Omega_{\rm GW}$ as a function of $f$ for some relevant values of the parameters and setting as an example $g_*=2\tt 10^2$: this is a reasonable value given that the SM sector has to be extended to satisfy the TAF requirement. In the same plot we also compare these theoretical findings with the experimental sensitivities of GW detectors (see Sec.~\ref{GW_det}).

Note that any cosmic source of GW background acts as an extra radiation component and, therefore, modifies the expansion rate of the Universe. This means that it is highly constrained by big-bang nucleosynthesis (BBN) measurements of primordial elements~\cite{Boyle:2007zx,Stewart:2007fu,Kohri:2018awv}.
 The measurement of the effective number of neutrino species $N_{\rm eff}$ and the observational abundance of dueterium and helium, gives us \oo
 the following bound~\cite{Maggiore:2018sht}:
 \be \int_{f_{\rm BBN}}^{f_{\rm UV}} \frac{df}{f} h^2 \Omega_{\rm GW}(f) <1.3 \tt 10^{-6}  \, \frac{N_{\rm eff}-3.046}{0.234},\label{BBNbound}\ee%  \be \Omega_{\rm GW} h^2 < \Omega_{\rm BBN} h^2 < 1.7 \times 10^{-6} \quad (95\%{\rm CL}),
%  \ee 
%In Ref.~\cite{Kohri:2018awv} they find 1.8 instead of 1.7. In~\cite{Stewart:2007fu} they have a different bound for the integral of $\Omega_{\rm GW} h^2$ (see their equation (15)).
where $f_{\rm BBN}\sim 10^{-11} $Hz~\cite{Maggiore:2018sht} (see e.g.~\cite{Boyle:2007zx,Maggiore:2018sht}) and $f_{\rm UV}$ is some UV cutoff, which in our case can be conservatively taken to be $\Lambda_G$, the scale above which gravity is softened. 
We have explicitly checked that this bound is satisfied 
%(for all numerical calculations performed in this work) 
by the theoretical curves in Figs.~\ref{Omega} 
 by carefully taking  into account  the integration in the left-hand-side of~(\ref{BBNbound}) and using $\Lambda_G\ll \bp$. Note, however, that the bound in~(\ref{BBNbound}) can also be applied to the non-integrated GW spectrum unless such  spectrum has a very narrow peak (see e.g.~\cite{Boyle:2007zx,Maggiore:2018sht}), which is not present  in any of our results. For simplicity we, therefore, show this bound in Fig.~\ref{Omega}, specifying that it applies to the integrated GW spectrum on the left hand side of~(\ref{BBNbound}). In doing so we use the reference value $N_{\rm eff} =3.046+0.234=3.28$, which corresponds to an experimental upper bound on $N_{\rm eff}$ at 95\% c.l.~\cite{Maggiore:2018sht}.

%experimental lines are those of [Bhupal Dev - Ferrer - Zhang - Zhang 1905.00891.pdf]

 Finally, in Fig.~\ref{Omegapeak} we provide the predictions of the model for the quantities $f_{\rm peak}$ and $h^2\Omega_{\rm peak}\equiv h^2\Omega_{\rm GW}(f_{\rm peak})$ for allowed values of the parameters of the axion sector ($f_a$ and $\bar g_s$). We considered values of $f_a$ starting from $10^9$~GeV (for which the axion contributes negligibly to DM, but is still compatible with astrophysical bounds) up to values such that the axion accounts for the whole DM. For each fixed value of $f_a$ the coupling $\bar g_s$ is varied above the minimal value discussed in Sec.~\ref{PQ-PT}. In Fig.~\ref{Omegapeak} the reference value $g_*=2\tt 10^2$ is again used. As we will see in the next section, these predictions  can be efficiently tested through GW observations.

%\iffalse
%\begin{figure}[t]
%\begin{center}
%\includegraphics[scale=0.6]{GW_TAF} 
%\end{center}
%	\caption{\em  The blue line gives $h^2\Omega_{\rm GW}(f)$ for $g_*=2\tt  10^2$ and $\{\bar g_s,f_a\} \approx  \{0.91,1.2\tt 10^9\,{\rm GeV}\}$.
%	%and $\{\bar g_s,f_a\} \approx  \{0.97,10^{11}\,{\rm GeV}\}$. 
%	The other colored lines correspond to projected sensitivities curves for various GW observatories as details in section \ref{GW_det}. 
%	\ag{BBN bound on this plot - Appendix C of Ref. 1804.08577. Do you have a bound on $N_{eff}$
%	from the extra dark photon present in the TAF axion model ?} \xxx{These bounds are satisfied as discussed in~\cite{Salvio:2020prd}.}
%	}
%\label{Omega}
%\end{figure}
%\fi

%\begin{figure}[H]
%\begin{center}
%\includegraphics[scale=0.5]{bht_1} 
%\end{center}
%	\caption{\em Same as the previous figure, for LISA.}	
%	\label{beta_H_T_LISA}
%\end{figure}

%    \vspace{0.5cm}
%    
%\subsection*{Acknowledgments}
%We thank ...
%
%\vspace{0.7cm}

\section{Gravitational wave detectors}
\label{GW_det}

%RGE runnings upto high-energy scales lies at the heart of models involving TAF principle and any prediction concerning such a scenario often involves heavy mass scales ($f_a$ here) is beyond the reach of any collider experiments. 

GWs serve as a probe for early universe cosmology, and is particularly important for the era prior to BBN. While inaccessible in collider or in other terrestrial experiments due to the high energy scales involved, the strong first-order phase transition associated with the PQ symmetry breaking induces a stochastic GW source and, therefore,  gives us a window to experimentally detect a PQ model.  Moreover, observing the particular GW spectrum predicted by the TAF axion model may serve as a first indirect verification of the TAF principle.

 \begin{figure}[t]
\begin{center}
\includegraphics[scale=0.55]{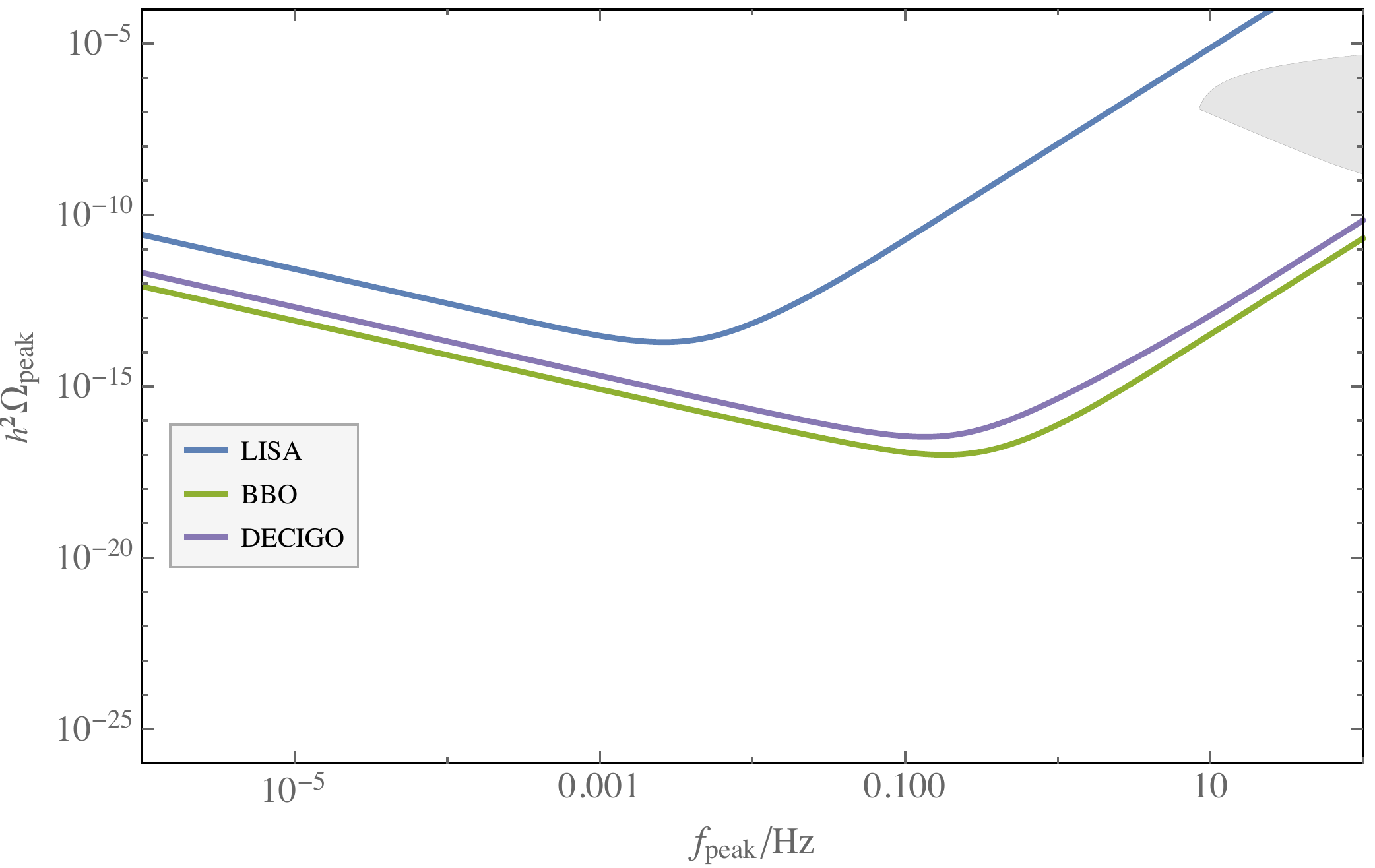}
\end{center}
	\caption{\em  Predictions of the model in the $(f_{\rm peak},h^2\Omega_{\rm peak})$ plane (grey region) by taking $f_a\geq 10^9~{\rm GeV}$ (compatibly with astrophysical observations) and $\bar g_s$ above the minimal values discussed in Sec.~\ref{PQ-PT}. Peak-integrated sensitivity curves for LISA, BBO and DECIGO are also shown.}. 
\label{Omegapeak}
\end{figure}

Although there may be other sources of stochastic GWs in the early universe like inflation (Refs.~\cite{Easther:2006gt,Turner:1996ck,Kuroyanagi:2018csn,Bartolo:2016ami}) or unidentified binary black hole mergers (Ref.~\cite{TheLIGOScientific:2016wyq}) the spectrum of GW radiation produced by phase transitions is generically different (see Ref.~\cite{Maggiore:2018sht,Mazumdar:2018dfl} for reviews). 
For instance, the expected stochastic background from compact binary coalescences, such as binary neutron stars or black holes, has a different dependence on frequency than the one in~(\ref{eq:gw_col}), 
%  In other astrophysical sources (for instance, the expected stochastic background from binary black hole mergers) the GW spectrum has a different dependence on frequency than the one in~(\ref{eq:gw_col}),   
  $\Omega_{\rm GW} \sim f^{2/3}$  \cite{Wu:2011ac}.  Various limits on astrophysical sources contributing as stochastic GW background exist~\cite{Abbott:2009ws}. In 2015, when the GW event (named GW150914) from binary back hole mergers was observed  \cite{Abbott:2016blz,TheLIGOScientific:2016wyq1} and its contribution reanalyzed as a source of stochastic background (with 90 \% C.L. statistical uncertainty, propagated from the local rate measurement, on the total background) the event was found to be in the frequency range of 0.01 Hz - 100 Hz, and of amplitude within the reach of advanced LIGO (see Ref.~\cite{TheLIGOScientific:2016wyq} for details). Such range overlaps with the one   of the PQ phase transition in the TAF axion model. This, however, is not a problem: due to its weaker frequency dependence, any \textit{multiple network} of GW detectors, like LIGO~\cite{TheLIGOScientific:2014jea,Harry:2010zz}, VIRGO~\cite{TheVirgo:2014hva}, GEO600~\cite{Luck:2010rt}, KAGRA~\cite{Somiya:2011np}, and LIGO-India~\cite{Unnikrishnan:2013qwa}, will be able to separate the astrophysical signal from other sources, like the one from a cosmological PQ phase transition or other events taking place after cosmic inflation. In the TAF axion scenarios, future GW detectors will be able to probe the model  as we will discuss in detail below.

All experiments to detect GWs has strain noise power spectrum
% ($\Omega_{\rm noise} = 2\pi^2/(3H_0^2f^3) S_{\rm noise}$)
 ($\Omega_{\rm noise} =2\pi^2 f^3S_{\rm noise}/(3H_0^2)$, where $H_0$ is the Hubble parameter in the present Universe, see e.g.~\cite{Schmitz:2020syl}). %\xxx{correct?! See Eq.~(A.24)  of 2002.04615}. 
The quantity $S_{\rm noise} (f)$ consists of intrinsic noise from the instrument as well as other astrophysical confusion noise $S_{\rm noise} (f) =S_{\rm ins} (f) + S_{\rm gcn} (f)$ (see Ref.~\cite{Alanne:2019bsm} for details) that varies from detector to detector.  This leads to a \textit{signal-to-noise ratio}
${\cal R}$ given by ~\cite{Allen:1997ad}
\begin{align}
\label{eq:snr}
{\cal R}^2 = N\,t_{\rm obs}
\int_{f_{\rm min}}^{f_{\rm max}}df
\left[\frac{\Omega_{\rm signal}\left(f\right)}{\Omega_{\rm noise}\left(f\right)}
\right]^2 \,,
\end{align}
%
%\xxx{In Eq. (2.6) of~\cite{Alanne:2019bsm} ${\cal R}^2$ contains $s\, {\rm Hz}$ at the denominator. Why did you drop it?} 
where $t_{\rm obs}$ denotes the experiment's observing time, $N=1$ ($N=2$)
for experiments that perform an auto (cross) correlation measurement of the stochastic gravitational wave background and 
$f_{\min}$ and $f_{\max}$ are the minimum and maximum frequency accessible to the detector,  respectively.
The GW spectrum can be detected by ground-based interferometers; these include advanced LIGO in Hanford and Livingston~\cite{Harry:2010zz,TheLIGOScientific:2014jea}, Cosmic Explorer (CE)~\cite{Evans:2016mbw,Reitze:2019iox}, 
Einstein Telescope (ET)~\cite{Punturo:2010zz, Hild:2010id, Sathyaprakash:2012jk}).
Moreover, additional information on the GW spectrum can be obtained through space-based interferometers
(BBO~\cite{Crowder:2005nr, Corbin:2005ny, Harry:2006fi}, DECIGO~\cite{Seto:2001qf, Kawamura:2006},
and LISA~\cite{Audley:2017drz}).
%
%For a $t_{\rm obs} = 4$~yr observing time,
%the interferometer experiments and the equivalent to  
%$t_{\rm obs} = 20$~yr running for the IPTA and SKA~\cite{Siemens:2013zla}. 
%\fh{We can remove this sentence: - - \ag{DONE!}} 
%A Fisher information analysis can marginally improve the SNR-based approach (see~\cite{Caldwell:2018giq}) we consider here. 

%\textcolor{red}{KAGRA \& CE if possible to extract from the paper.}

%\fh{I think we have already enough constraint curves in the plot. It is better not to consider them. KAGRA limit is higher than LIGO and CE limit is in the similar range of ET.
% - -\ag{OK}}
%
%In future work, one might consider refining our analysis by means of a Fisher information analysis.
%
%We, however, expect that such an analysis would only marginally improve over our SNR-based approach (see~\cite{Caldwell:2018giq}, which finds no difference between both approaches in the specific case of LISA).

%%%%%%%%%%%%%%%%%%%%%%%%%%%%%%%%%%%%%%%%%%%%%%%%%%%%%%%%%%%%%%%%%%%%%%%%%%%%%%%%%%%%%%%%%%%%%%%%%%%%

%In principle, it is also necessary to account for 
There are astrophysical sources contributing as confusion noise (see e.g.~\cite{Cornish:2018dyw}) but its impact becomes insignificant
with time as one is able to subtract this due to information from the
individual foreground sources, which increases in number. However, we consider 
no foreground contamination, therefore, the results can be treated as an upper limit of the experimental reach possible in future.

In Fig.~\ref{Omega}, we depict the predicted GW spectra for some benchmark points 
 along with the projected sensitivities of current and future interferometer experiments~\cite{Corbin:2005ny,Punturo:2010zz,Evans:2016mbw,Audley:2017drz,Musha:2017usi,Cornish:2018dyw,LIGOScientific:2019vkc,Thrane:2013oya,Moore:2014lga}.  %Blue, red, purple, green and orange curves from left to right correspond to points A, B, C, D and E respectively.  
 The benchmark points in Fig.~\ref{Omega} include physically interesting cases with $f_a\sim10^9~$GeV and $f_a\sim 10^{11}~$GeV for different values of $\bar g_s$.
Each GW projected sensitivity is denoted with corresponding legends. % \xxx{sentence commented} 
%Curves are drawn following peak-integrated sensitivity curves\footnote{We have also checked the same with power-law-integrated sensitivity curves using gwplotter \cite{Thrane:2013oya,Moore:2014lga}.} 
% from Refs.~\cite{Alanne:2019bsm,Schmitz:2020syl}.
   As clear from Fig.~\ref{Omega}, Advanced LIGO, ET, CE, DECIGO and BBO  will be potentially able to detect GWs produced by the PQ phase transition in the TAF axion model.

Fig.~\ref{beta_H_T} provides sensitivity plots in the ($ T_n, \beta/H $) plane for various experiments and compare them with the theoretical curves from the TAF axion model. We considered the physically interesting cases $f_a=10^{11}~$GeV and $f_a=10^9~$GeV.
 Fig.~\ref{beta_H_T} shows that ET, CE, BBO and DECIGO will be able to test this scenario through measurements of $ T_n$ and $\beta/H$.
 
In Fig.~\ref{Omegapeak} we compare the predictions of the TAF axion model in the $(f_{\rm peak},h^2\Omega_{\rm peak})$ plane with the peak-integrated sensitivity curves for LISA, BBO and DECIGO. These curves are drawn following the method explained in Ref.~\cite{Schmitz:2020syl}. Looking at  Fig.~\ref{Omegapeak} one can clearly see that BBO and DECIGO have the potential to test the allowed parameter space of the TAF axion model.

Finally, we conclude this section by noting that the possible ({\it if any})  strong first-order phase transitions due to the TAF SM sector might also produce detectable GWs. The combination of various planned experiments, which have different sensitivity ranges, can help in distinguishing the PQ phase transition from the possible phase transitions due to TAF SM sector. We do not enter the discussion of these SM phase transitions as highly dependent on the specific SM sector one considers. We leave such specific analysis  to future work.

%
%
%In Fig.~\ref{Omega}, we depict the predicted GW spectra for our benchmark points 
% along with the projected sensitivities of current and future interferometer experiments.  %Blue, red, purple, green and orange curves from left to right correspond to points A, B, C, D and E respectively.  
%Each GW projected sensitivity is denoted with corresponding legends. Curves are drawn following peak-integrated sensitivity curves  from Refs.~\cite{Alanne:2019bsm,Schmitz:2020syl}. We have also checked the same with power-law-integrated sensitivity curves using gwplotter \cite{Thrane:2013oya,Moore:2014lga}. Advanced LIGO, ET, CE, DECIGO, BBO \xxx{...} will be able to detect GWs produced by the PQ phase transition in the fundamental QCD axion model \xxx{...}
%
%Fig.~\ref{beta_H_T} provides sensitivity plots in the ($ T_n, \beta/H $) plane \xxx{...}

%\ag{Which value of bubble wall velocity is used (usually 0.6 is taken) ? Can we also check with 0(0.1) deviations from this value as these bubble wall velocities are from lattice simulations, and as the simulations get improved in future the exact values might change ? However significant deviations are not expected.}

%\medskip

\section{Conclusions}\label{Conclusions}
  
 The detection of GWs and many upcoming GWs detectors have recently reinforced  the interest in phase transitions predicted by particle physics models.  In this paper we have studied the PQ phase transition and the corresponding spectrum of GWs in a QCD axion model  where all couplings flow to zero in the infinite  energy limit (TAF property) and the PQ  symmetry breaking scale $f_a$  is generated quantum mechanically through the CW mechanism.  This fundamental (i.e. UV complete) model features an extra gauge group SU(2)$_a$, which is spontaneously broken to an Abelian U(1)$_a$ subgroup. The low-energy spectrum, therefore, includes a dark photon, which has previously been shown to be compatible with current bounds from particle physics experiments and cosmology~\cite{Salvio:2020prd}.  This TAF QCD axion model is highly predictive; indeed, the axion sector has only one independent dimensionful quantity, $f_a$, and one independent dimensionless parameter, $\bar g_s$. Therefore, the masses of the particles in the axion sector are all predicted in terms of these two parameters.
  
We have found  that this model features a  first-order PQ %
phase transition, which is very strong. The presence of only  few adjustable parameters results in interesting predictions regarding the main quantities associated with the phase transition, $T_n$, $\beta/H_I$, etc, mainly summarized in the left plot of Fig.~\ref{Tnvsgs} and in Fig.~\ref{beta_H_T}. We have shown that, like in previous effective axion models with CW PQ symmetry breaking, the phase transition is characterized by a period of strong supercooling, $T_n \ll T_c$, when the universe inflated. Thanks to this period the monopole density associated with the breaking SU(2)$_a\to\, \,$U(1)$_a$ is efficiently diluted. Reheating then generically occurs via the unavoidable couplings between the SM and the axion sector due to gluons, which guarantee a rather large reheating temperature. For $\bar g_s$ of order 1, the model  predicts  values of $T_n$ and $\beta/H_I$ that are within the reach of future GW detectors, such as ET, CE, DECIGO and BBO (see Fig.~\ref{beta_H_T}). 

The key theoretical tool, which we have used to obtain these results regarding the first-order phase transition, is the calculation of the bounce solutions associated with the tunnelling from the PQ symmetric configuration to the PQ breaking vacuum. Within this formalism   we have also checked that the phase transition is mainly due to thermal effects rather than quantum effects: the action of the O(3)-symmetric bounce divided by the temperature is always much smaller than the action  of the O(4)-symmetric one.   
  
    Finally, the predictivity of the model also interestingly leads to %specific features 
a rigid dependence of the GWs spectrum produced by the PQ phase transition on $f_a$ and $\bar g_s$. 
  We have compared this theoretical spectrum with the sensitivities of several future detectors such as ET, CE, DECIGO, BBO and advanced LIGO (see Fig.~\ref{Omega} and~\ref{Omegapeak}), finding conclusively that these experiments will be able to test the fundamental QCD axion model.

We believe that the precision that GW astronomy promises due to the planned worldwide network of GW detectors can make the dream of testing high-scale and fundamental BSM scenarios of UV-completion a reality in near future.
  
     \vspace{0.4cm}
    
\subsection*{Acknowledgments}
We thank M.~Bianchi, A.~Dasgupta, R.~Frezzotti, S.~Hajkarim, N.~Tantalo and G.~White for useful discussions.

\end{document}